\documentclass[10pt,twocolumn,letterpaper]{article}

\usepackage{cvpr}              

\usepackage{graphicx}
\usepackage{amsmath}
\usepackage{amssymb}
\usepackage{booktabs}
\usepackage{float}

\usepackage{siunitx}
\def\degree{${}^{\circ}$}
\usepackage{bm}
\usepackage[dvipsnames]{xcolor}

\usepackage{algorithm}
\usepackage[noend]{algpseudocode}
\usepackage{multirow}
\usepackage{booktabs}

\usepackage{xcolor}
\definecolor{Cyan}{cmyk}{1,0,0,0}

\newcommand{\hyh}[1]{{\color{black}#1}}
\newcommand{\review}[1]{{\color{black}#1}}

\usepackage[pagebackref,breaklinks,colorlinks]{hyperref}

\usepackage[capitalize]{cleveref}
\crefname{section}{Sec.}{Secs.}
\Crefname{section}{Section}{Sections}
\Crefname{table}{Table}{Tables}
\crefname{table}{Tab.}{Tabs.}

\def\cvprPaperID{133} 
\def\confName{CVPR}
\def\confYear{2022}

\begin{document}

\title{StylizedNeRF: Consistent 3D Scene Stylization as Stylized NeRF via 2D-3D Mutual Learning}

\author{
\textbf{Yi-Hua Huang}\textsuperscript{1,2}
\quad 
\textbf{Yue He}\textsuperscript{1,2}
\quad 
\textbf{Yu-Jie Yuan}\textsuperscript{1,2}
\quad 
\textbf{Yu-Kun Lai}\textsuperscript{3}
\quad 
\textbf{Lin Gao}\textsuperscript{1,2*} \\
\textsuperscript{1}Beijing Key Laboratory of Mobile Computing and Pervasive Device,\\ Institute of Computing Technology, Chinese Academy of Sciences\\
\textsuperscript{2}School of Computer and Control Engineering, University of Chinese Academy of Sciences \\
\textsuperscript{3}School of Computer Science \& Informatics, Cardiff University\\
{\tt\small \{huangyihua20g, heyue19s, yuanyujie, gaolin\}@ict.ac.cn} \quad
{\tt\small LaiY4@cardiff.ac.uk}
}

\twocolumn[{%
\renewcommand\twocolumn[1][]{#1}%

\maketitle

\vspace{-9mm}
\begin{center}
    \centering
    \captionsetup{type=figure}
    \includegraphics[width=0.95\linewidth]{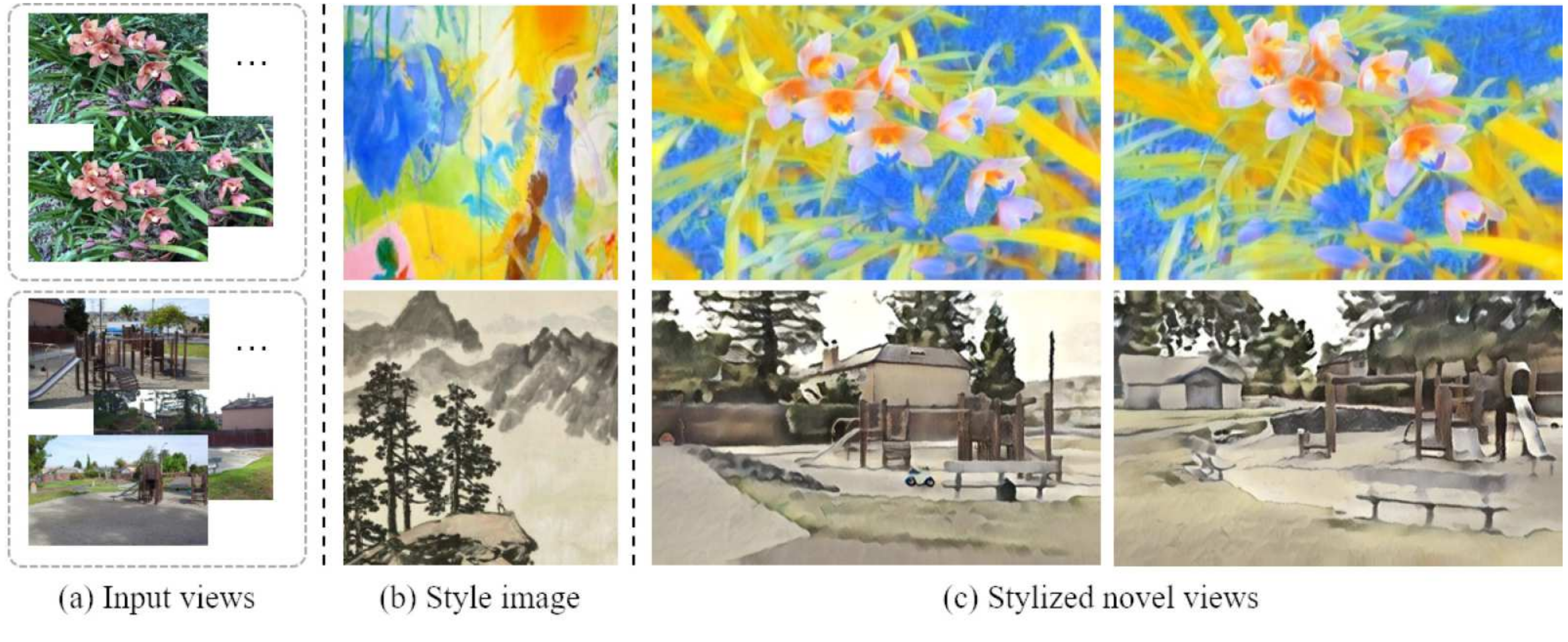}
    \vspace{-4mm}
    \captionof{figure}{\textbf{Results of consistent 3D stylization by our method}. Given a set of real photographs (a) and a style image (b), our model is capable of generating stylized novel views (c), which are consistent in 3D space by learning a stylized NeRF.
    }
\label{fig:teaser}
\end{center}
}]

\if TT\insert\footins{\noindent\footnotesize{
*Corresponding Author is Lin Gao (gaolin@ict.ac.cn).}}\fi

\begin{abstract}
3D scene stylization aims at generating stylized images of the scene from arbitrary novel views following a given set of style examples, while ensuring consistency when rendered from different views. Directly applying methods for image or video stylization to 3D scenes cannot achieve such consistency. Thanks to recently proposed neural radiance fields (NeRF), we are able to represent a 3D scene in a consistent way. Consistent 3D scene stylization can be effectively achieved by stylizing the corresponding NeRF. However, there is a significant domain gap between style examples which are 2D images and NeRF which is an implicit volumetric representation. To address this problem, we propose a novel mutual learning framework for 3D scene stylization that combines a 2D image stylization network and NeRF to fuse the stylization ability of 2D stylization network with the 3D consistency of NeRF. We first pre-train a standard NeRF of the 3D scene to be stylized and replace its color prediction module with a style network to obtain a stylized NeRF. It is followed by distilling the prior knowledge of spatial consistency from NeRF to the 2D stylization network through an introduced consistency loss. We also introduce a mimic loss to supervise the mutual learning of the NeRF style module and fine-tune the 2D stylization decoder. In order to further make our model handle ambiguities of 2D stylization results, we introduce learnable latent codes that obey the probability distributions conditioned on the style. They are attached to training samples as conditional inputs to better learn the style module in our novel stylized NeRF. Experimental results demonstrate that our method is superior to existing approaches in both visual quality and long-range consistency.
\end{abstract}

\begin{figure*}[htbp]
	\centering
	\includegraphics[width=0.95\linewidth]{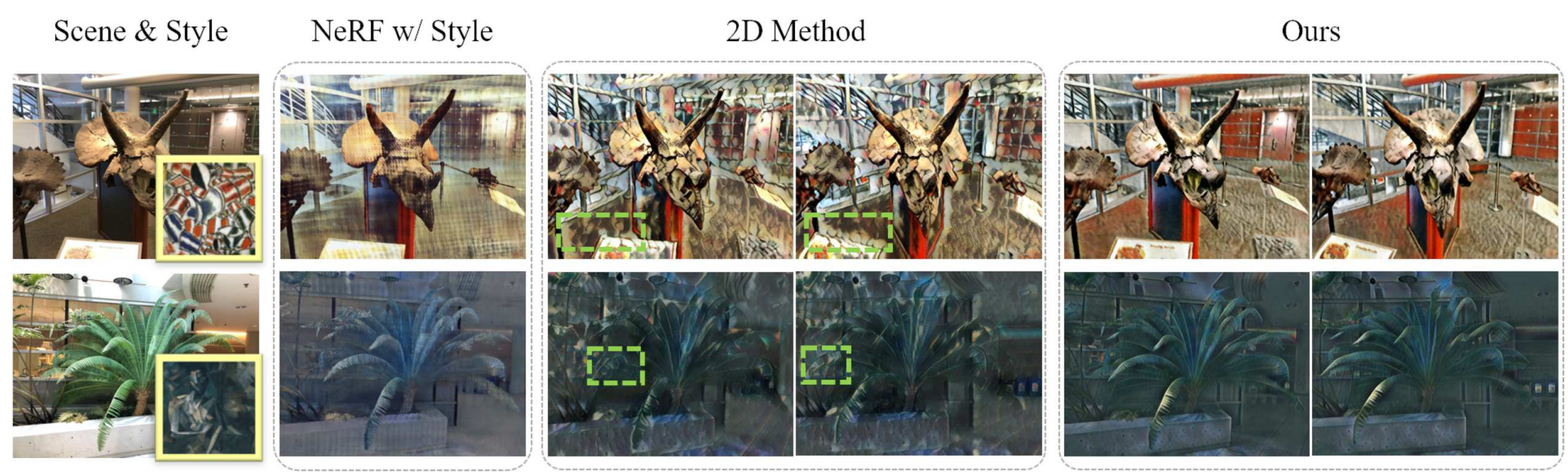}
	\caption{\textbf{Motivation on our mutual learning scheme.} Only training a stylized NeRF with style and content losses on small training patches will lead to poorly maintained content and unsatisfactory transfer of style (NeRF w/ Style). Directly applying a 2D image stylization method (AdaIN is used in this example) on results of NeRF will cause inconsistency when rendered from different views (2D Method). Our method of mutual learning the stylized NeRF and 2D stylization method produces results with better style and consistency quality.}
\label{fig:motivation}
	\vspace{-5mm}
\end{figure*}

\section{Introduction}
\label{sec:intro}

Controlling the appearance of complex 3D real scenes has attracted increasing attention in recent years. Numerous works have made great effort to this task, such as texture synthesis~\cite{xiang2021neutex,kanazawa2018learning,gao2020tmnet} and semantic view synthesis~\cite{habtegebrial2020generative,huang2020semantic}.
In this paper, we focus on the problem of stylizing complex 3D real scenes, which is useful for applications such as virtual reality and augmented reality.
Thanks to the recently advanced 3D representation methods, complex 3D scenes can be represented as point clouds with appearance features~\cite{riegler2021stable} or implicit fields by deep neural networks such as neural radiance fields (NeRF)~\cite{mildenhall2020nerf,zhang2020nerf++}. Compared with point clouds, NeRF can be more reliably obtained from multi-view images, and is continuous in 3D spaces, making learning easier.

In this paper, we aim to stylize a 3D scene following a given set of style examples. This allows generating stylized images of the scene from arbitrary novel views, while making sure rendered images from different views are consistent. To ensure consistency, we formulate the problem as stylizing a NeRF~\cite{mildenhall2020nerf} with a given set of style images. Some examples of our NeRF stylization method are presented in Fig.~\ref{fig:teaser}.

However, there are two challenges to leverage NeRF as the representation of a complex 3D scene in the task of stylization.
Firstly, NeRF needs to query hundreds of sample points along the ray to render a single pixel. The memory limitation makes it intractable to render the whole image or even a big enough patch at one time which is important for calculating content and style losses~\cite{johnson2016perceptual}. 
\review{Therefore, straightforwardly training a stylized NeRF with perceptual style and content losses on small training patches (32$\times$32 for a single RTX 2080Ti GPU) leads to poor stylization results, as shown in Fig.~\ref{fig:motivation}.}
Secondly, directly adopting state-of-the-art image stylization methods to stylize rendered images from NeRF will generate inconsistent results across different views~\cite{huang2021learning,chiang2022stylizing}. This is because these stylization methods lack 3D information. Taking a representative Adaptive Instance Normalization (AdaIN)~\cite{huang2017arbitrary} method for example, its results can be seen in the third and fourth columns in Fig.~\ref{fig:motivation}. On the other hand, training a NeRF with inconsistent 2D stylized images will cause blurriness in results, which will be further illustrated in Sec.~\ref{sec:experiment}. 

In order to tackle the problems mentioned above, we propose a novel mutual learning framework~\cite{zhang2018deep} between NeRF~\cite{mildenhall2020nerf} and a 2D image stylization method. 
An ordinary NeRF network 
is first trained to model the opacity field of the scene. 
\hyh{The opacity field of NeRF has the inherence of geometric consistency and can estimate the 3D coordinates of the rendered pixels, which is distilled to the 2D stylization method through a consistency loss at a pre-training stage.
To represent the stylized scene, we replace the module that predicts color in NeRF with a style module (referred to as stylized NeRF). We then co-train the novel stylized NeRF network (with density prediction fixed) with the pre-trained 2D stylization network for fine-tuning collaboratively. A mimic loss is introduced to align the outputs of the stylized NeRF and the 2D method, aiming to share the stylization knowledge of the 2D method and inherent geometry consistency of NeRF to update networks. However, the 2D stylization method cannot guarantee strict consistency, which leads to ambiguities when transferring a given style among multi-view frames of a certain 3D scene, resulting in blurry results of the stylized NeRF.  }

Inspired by NeRF-W~\cite{martin2021nerf}, 
our style module takes learnable latent codes as conditioned inputs to handle the ambiguities of the 2D stylized results. Unlike NeRF-W, we build a novel probability model of the latent codes conditional to styles, which enables our model to handle the inconsistency of 2D results and meanwhile to stylize the scene conditionally.
We first extract style features of style images with VGG~\cite{simonyan2014very}, which are defined as the mean and variance of feature maps along the spatial dimensions~\cite{huang2017arbitrary}. We then encode the style features into latent distributions using a pre-trained variational autoencoder (VAE)~\cite{kingma2013auto}. The encoded distributions are conditioned to the encoding style features. 
\hyh{Since the inconsistent stylized results generated from the 2D network can be considered to be different samples obeying the distributions conditional to styles, we parameterize 2D stylization results as latent codes obeying the distributions encoded by the corresponding styles.} A minus log-likelihood of latent codes is then applied to constrain the conditional probability modeling of latent codes and further ensure the robustness of conditional stylization.

Our main technical contributions are as follows:
\begin{itemize}
    \item We propose a novel stylized NeRF approach for stylizing 3D scenes with given style images, outperforming existing methods in terms of visual quality and 3D consistency.
    \item We propose a mutual learning strategy for the stylized NeRF and 2D stylization method, leveraging the stylization capability of 2D method and geometry consistency of NeRF.
    \item A conditional probability modeling for learnable latent codes is proposed to handle the ambiguities of 2D stylized results while enabling conditional stylization. 
\end{itemize}

\section{Related Work}
\label{sec:relate}

\noindent \textbf{Novel View Synthesis.} Various methods have been proposed to synthesize novel views of a scene with a given set of photographs. Traditional light field techniques~\cite{gortler1996lumigraph,davis2012unstructured,levoy1996light} interpolate the dense input images as 2D slices of a 4D function to render novel views.
Multi Plane Image (MPI)~\cite{srinivasan2019pushing,tucker2020single,wizadwongsa2021nex,zhou2018stereo,mildenhall2019local} uses RGBD layers of different depths to represent the light field of the scene. Novel views can be rendered by warping the layers and compositing them to an image.
Some works utilize explicit 3D proxies, such as meshes~\cite{waechter2014let,buehler2001unstructured,debevec1996modeling,wood2000surface}, point clouds~\cite{niklaus20193d,wiles2020synsin,meshry2019neural,aliev2020neural} or voxels~\cite{seitz1999photorealistic,flynn2019deepview,lombardi2019neural,sitzmann2019deepvoxels} to reconstruct the scene. Based on the 3D proxy, existing work combines the geometry with methods representing appearance like colors~\cite{waechter2014let,seitz1999photorealistic}, texture mapping~\cite{debevec1996modeling}, light fields~\cite{buehler2001unstructured,wood2000surface} or deep networks for neural rendering~\cite{flynn2019deepview,lombardi2019neural,sitzmann2019deepvoxels,niklaus20193d,wiles2020synsin,meshry2019neural,aliev2020neural}.
Recently proposed image based works~\cite{riegler2020free,riegler2021stable} estimate the 3D proxy in the form of a point cloud 
and neurally render the novel views.

A recent trend of continual neural representations is to replace discrete 3D proxy representations 
with MLPs mapping 3D coordinates to the property of corresponding locations. Implicit functions~\cite{niemeyer2020differentiable,yariv2020multiview} model the implicit surfaces of the scenes. NeRF~\cite{mildenhall2020nerf} further models the radiance fields of the scene as particles emitting and blocking lights. The following works extend NeRF to octree structure~\cite{liu2020neural}, unbounded scenes~\cite{zhang2020nerf++}, reflectance decomposition~\cite{boss2021nerd} and uncontrolled real-world images~\cite{martin2021nerf}. Please refer to \cite{tewari2021advances} for the recent developments in neural rendering.

\noindent \textbf{Style Transfer Methods.}
Style transfer is a long-standing research topic in computer vision. 
Gatys~\etal~\cite{Gatys2015TextureSU} is a pioneering work in this field focusing on an optimization-based scheme. 
For faster stylization, follow-up works turn to leverage feed-forward neural networks, such as Avatar \cite{sheng2018avatar}, and AdaIN \cite {huang2017arbitrary}.
Li~\etal~\cite{Univeral_style_transfer} proposed a method that embeds whitening and coloring transformation (WCT) to generate high resolution stylized images. Methods based on it sprung up like PhotoWCT \cite{li2018closed}, WCT$^2$\cite{yoo2019photorealistic}.
Instead of optimizing on pixels, \cite{kotovenko2021rethinking} optimizes directly on parameterized brushstrokes. The effect is stunning when the style image is full of delicate strokes. 

Video stylization is another topic in this field since it demands consistency between adjacent frames to ensure the stylized video is free of flickering. Most methods are based on optic flow or simply add a temporal constraint to existing methods \cite{gao2020fast,chen2017coherent}. The work \cite{deng:2020:arbitrary} aligns cross-domain features with input videos and achieves coherent results and \cite{wang2020consistent} dynamically adjusts inter-channel distributions based on relaxation and regularization. 

These 2D-based methods lack a spatial consistency constraint and 3D scene perception, thus do not have the ability to maintain long-term consistency in our task. Huang~\etal~\cite{huang2021learning} extend stylization to 3D scenes where a 3D scene is represented as a point cloud. It ensures consistency even in 360\degree unbounded scenes. 
\cite{chiang2022stylizing} firstly introduces NeRF to 3D scene stylization. The NeRF of ~\cite{chiang2022stylizing} is trained with style and content losses by sub-sampling patches to cope with the high complexity. As a result, the method tends to lose fine details. \hyh{Some other works~\cite{sumner2004deformation,gao2018automatic,zheng2021weakly} focus on transferring the pose style of 3D models or 2D human skeletons.} 

\begin{figure*}[htbp]
	\centering
	\includegraphics[width=0.85\linewidth]{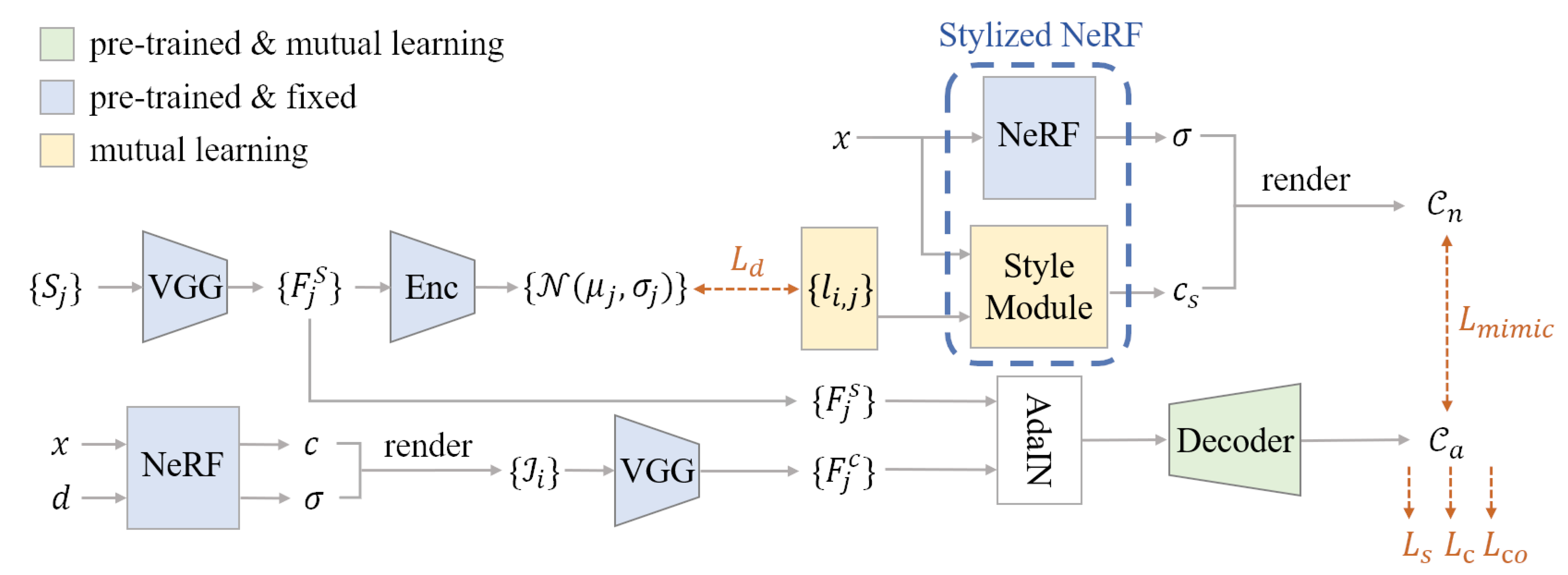}
	\caption{\textbf{The architecture of our model for mutual learning.} Our approach uses a pre-trained and fixed NeRF to render a number of views $\{\mathcal{I}_i\}$ as augmented data for mutual learning. The style features $\{F_j^s\}$ of style images $\{\mathcal{S}_j\}$ extracted by VGG are embedded to latent distributions through the encoder of a pre-trained VAE. The extracted content features $\{F_i^c\}$ together with the style features $\{F_j^s\}$ are fed into an AdaIN layer and a decoder to obtain the stylized colors $\mathcal{C}_a$. On the other side the style module takes the input of learnable latent codes $\{l_{i,j}\}$ and coordinates $x$ to predict the stylized radiance color $c_s$, which forms the stylized NeRF. By composition of the sampling points along the ray with original opacity $\sigma$, the rendered stylized color $\mathcal{C}_n$ can be obtained. The objective functions $L_d$, $L_{mimic}$, $L_{s}$, $L_{c}$ and $L_{co}$ are used for mutual learning optimization (see the text for details).}
\label{fig:pipeline}
	\vspace{-4mm}
\end{figure*}

\section{Preliminaries}

NeRF~\cite{mildenhall2020nerf} uses MLPs to model a scene as a continuous volumetric field of opacity and radiance. The MLPs take 3D position $\bm{x}\in\mathbb{R}^3$ and viewing direction $\bm{d}\in\mathbb{R}^2$ as input and predict opacity $\sigma(\bm{x})\in\mathbb{R}^+$ and radiance color $c(\bm{x},\bm{d})$. During rendering process, a ray $\bm{r}(t)=\bm{o}+t\bm{d}$ is cast from the camera's center $\bm{o}\in\mathbb{R}^3$ along the direction $\bm{d}$ passing through the pixel. The color of the pixel is determined by the integral: 

\begin{equation}
\label{eq:render}
    C(\bm{r})=\int_{t=0}^\infty \sigma(\bm{o}+t\bm{d}) c(\bm{o}+t\bm{d},\bm{d}) e^{-\int_{s=0}^t \sigma(\bm{o}+s\bm{d})ds} dt
\end{equation}

To facilitate the fitting capability of the model, NeRF uses positional encoding $\gamma(\cdot)$ to map inputs of the network $\bm{x}$ and $\bm{d}$ to their Fourier features~\cite{tancik2020fourier} containing signals of muti-scale frequencies:

\begin{equation}
    \gamma(x) = [\sin(x), \cos(x), ..., \sin(2^{L-1}x), \cos(2^{L-1}x)]^T
\end{equation}
where $L$ is a hyper parameter controlling the spectral bandwidth.

\section{Method}

We now illustrate our framework for stylizing a 3D scene with given style images. Given a collection of images from a scene with corresponding camera parameters, our goal is to generate stylized images following the given style from specified novel views while keeping geometry consistency. 
\hyh{To achieve this, we propose a mutual learning scheme to optimize the newly introduced stylized NeRF and the 2D stylization network mutually through consistency and mimic losses.}
Even though the mutually learned stylized NeRF is inherent consistent, 2D stylization network cannot guarantee  strict consistency in results, which will still cause blurriness in the results of the stylized NeRF. Therefore, we propose to view the inconsistent 2D stylization results as different samples obeying the distributions conditioned by the style and introduce latent codes that obey such conditioned distributions to handle the inconsistency. 
\hyh{We model learnable latent codes with conditional probability through a minus log likelihood loss.}
In the next, we will first introduce the 2D stylization network we adopt in Sec.~\ref{sec:2d_style}, and then discuss our stylized NeRF in Sec.~\ref{sec:3d_style}. Finally, we describe how we build the mutual learning framework based on the two networks in Sec.~\ref{sec:mutual}.

\subsection{2D Stylization Network} \label{sec:2d_style}
We adopt AdaIN~\cite{huang2017arbitrary} as our 2D stylization method, which consists of a VGG~\cite{simonyan2014very} encoder, an adaptive instance normalization layer, and a CNN-based decoder. It should be noted that AdaIN is a representative method but may be replaced with other advanced image stylization methods. 
The feature maps are firstly extracted by the encoder from given input style and content images~\cite{gatys2015neural} and then the adaptive instance normalization layer aligns the mean and variance of the content feature maps to the style feature maps. Finally the decoder decodes the aligned feature maps and generates output results with the target style.
During the training process, only the decoder of AdaIN is learnable. 
\hyh{We pre-train the decoder with distilled 3D consistency knowledge from NeRF through a consistency loss $L_{co}$ in addition to style and content losses.}
$L_{co}$ is calculated by warping stylized images from different views to a fixed one according to the geometry prior from NeRF:
\begin{equation}
\label{eq:consistency}
    L_{co} = ||O_{i,s} - M_{i,j}W_{i,j}(O_{j,s})||^2
\end{equation}
where $O_{i,s}$ denotes the stylized result of view $i$ and style $s$.

$W_{i,j}$ denotes the warping operation from view $j$ to view $i$ according to the depth estimated by NeRF and $M_{i,j}$ denotes the mask of warping and occlusion.

\subsection{Stylized NeRF} \label{sec:3d_style}
An ordinary NeRF~\cite{mildenhall2020nerf} 
is trained to model the opacity field $\sigma(x)$ and original radiance color field $c_o(x,d)$, which is fixed in the following mutual learning process. To enable the stylization capability of NeRF, an MLP network is added as a style module to NeRF in place of the original color module, modeling the stylized radiance color of the scene. When querying the stylized radiance color of the scene in the training stage, the module takes the input of learnable latent codes in addition to position coordinates. Unlike the latents in NeRF-W that models random appearance and transients of the scene, the latent codes here both learn the style and ambiguities of the 2D stylization results, \hyh{avoiding blurriness in results of the stylized NeRF and enabling it to conditionally stylize the scene.}
 The stylized results of 2D methods on different views with the specified style can be regarded as samples of a conditional distribution. Those samples are different due to their inconsistency. We parameterize the conditional distributions of 2D stylized results through a pre-trained VAE~\cite{kingma2013auto}. The VAE encodes style features $\{F_j^s\}$ extracted by VGG as Gaussian distributions $\{\mathcal{N}(\mu_j,\sigma_j)\}$. The conditional distributions of 2D stylized results are parameterized as the embedded Gaussian distributions, which is conditioned with style features. For the 2D stylized result of the $i$-{th} view and $j$-{th} style, a latent code $l_{i,j}$ initialized by sampling on $\mathcal{N}(\mu_j, \sigma_j)$ is assigned to it. The latent codes are optimized during the mutual learning process. To constrain the latent codes $l_{i,j}$ to obey the distributions $\{\mathcal{N}(\mu_j,\sigma_j)\}$, a minus log likelihood loss $L_d$ is used: 

\begin{equation}
    L_d(l_{i,j}) = \frac{(l_{i,j} - \mu_j)^2}{2\pi\sigma_j^2}
\end{equation}
where $i$ and $j$ are the indices of the training view and style image, respectively. $\mu_j$ and $\sigma_j$ denote the mean and variance of the distribution embedded by the $j$-{th} style image. Hence we parameterize the conditional distributions of 2D stylized results by constraining the learnable latent codes to obey the distributions conditioned on styles. At inference time, the mean $\mu$ of embedded distribution is used as input to stylize the scene. The loss $L_d$ constrains the latent codes to obtain better clustering and generalization, thus leads to better results, as we will later demonstrate in Fig.~\ref{fig:logp}.
 
\begin{figure*}[htbp]
	\centering
	\includegraphics[width=1.\linewidth]{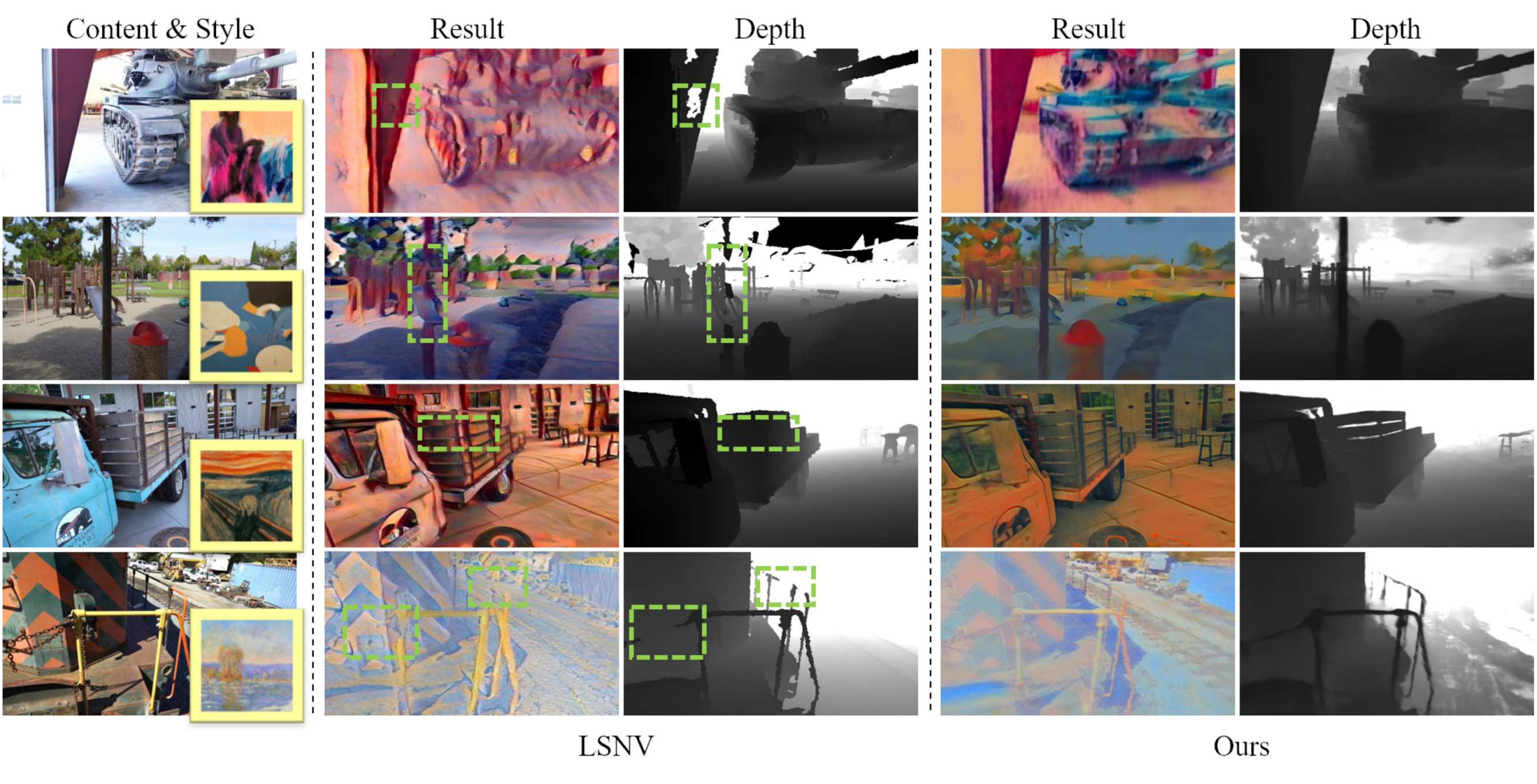}
	\caption{\textbf{Qualitative comparisons to LSNV.} We compare the stylized results of 4 scenes in Tanks and Temples dataset~\cite{knapitsch2017tanks}. Our method stylize scenes with more precise geometry and competitive stylization quality.}
\label{fig:vs_lsnv}
	\vspace{-4mm}
\end{figure*}

The style module takes in the embedded $\mu$ and 3D position coordinates to obtain the stylized color $c_s(x,l)$. The style module along with the opacity prediction module of the pre-trained NeRF forms our stylized NeRF.  The rendering procedure follows Eq.~\ref{eq:render}, which uses the original opacity field.

\subsection{Mutual Learning} \label{sec:mutual}
\hyh{The mutual learning starts with distilling spatial consistency prior knowledge from NeRF to the 2D stylization network through $L_{co}$ as described in ~\ref{sec:2d_style}.}
It is followed by the collaborative training of the learnable style module, a pre-trained decoder of AdaIN for fine-tuning and latent codes. To augment the training dataset, a series of views $\{\mathcal{I}_i\}$ are rendered by the ordinary NeRF as training data. We denote the style images as $\{\mathcal{S}_j\}$. A training view and a given style together form a training instance, to which a latent code $l_{i,j}$ described in Sec.~\ref{sec:3d_style} is assigned. Using the original opacity, the image of the stylized NeRF is rendered through sampling points along the ray and approximating Eq.~\ref{eq:render} by numerical quadrature, as discussed in~\cite{max1995optical}:

\begin{equation}
\begin{gathered}
    \mathcal{C}_{n}(r,l) = \sum\limits_{k=1}^{K}T_k(1-\exp(-\sigma_k\delta_k))c_s(r_k, l), \\ \text{where } T_k=\exp(-\sum\limits_{k'=1}^{k-1}\sigma_{k'}\delta_{k'})
\end{gathered}
\end{equation}
where $\mathcal{C}_n(r,l)$ is the predicted stylized color of the pixel $r$ and $\delta_k$ is the Euler distance between the $k$-{th} and $(k+1)$-{th} sample points. The mimic loss is defined as the L2 distance between the stylization result $\mathcal{C}_n(r_i,l_{i,j})$ from NeRF and $\mathcal{C}_a(\mathcal{I}_i, \mathcal{S}_j)_{r_i}$ from the 2D stylization method:

\begin{equation}
    L_{mimic} = \sum\limits_{i,j,r}||\mathcal{C}_n(r_i,l_{i,j}) - \mathcal{C}_a(\mathcal{I}_i, \mathcal{S}_j)_{r_i}||^2
\end{equation}
 The mimic loss is introduced to best exchange the knowledge of different strengths between the NeRF and 2D stylization method. The perceptual content loss $L_c(\mathcal{C}_a(\mathcal{I}_i, \mathcal{S}_j), \mathcal{I}_i)$ and style loss $L_s(\mathcal{C}_a(\mathcal{I}_i, \mathcal{S}_j), \mathcal{S}_j)$~\cite{huang2017arbitrary} are determined by the results of the decoder $\mathcal{C}_a(\mathcal{I}_i, \mathcal{S}_j)$, which allows larger patches within limited GPU memory. The objective function of the mutual learning process for the NeRF style module and latent codes is:

\begin{equation}
     L_{N} = L_{mimic} + \lambda_d L_d
\end{equation}
The objective function for fine-tuning the 2D stylization decoder can be written as
 \begin{equation}
     L_{C} = \lambda_mL_{mimic} + \lambda_s L_s + L_c
 \end{equation}
 where $\lambda_p$, $\lambda_c$, $\lambda_s$ are hyper parameters controlling the impact of terms.

\begin{figure*}[htbp]
	\centering
	\includegraphics[width=1.\linewidth]{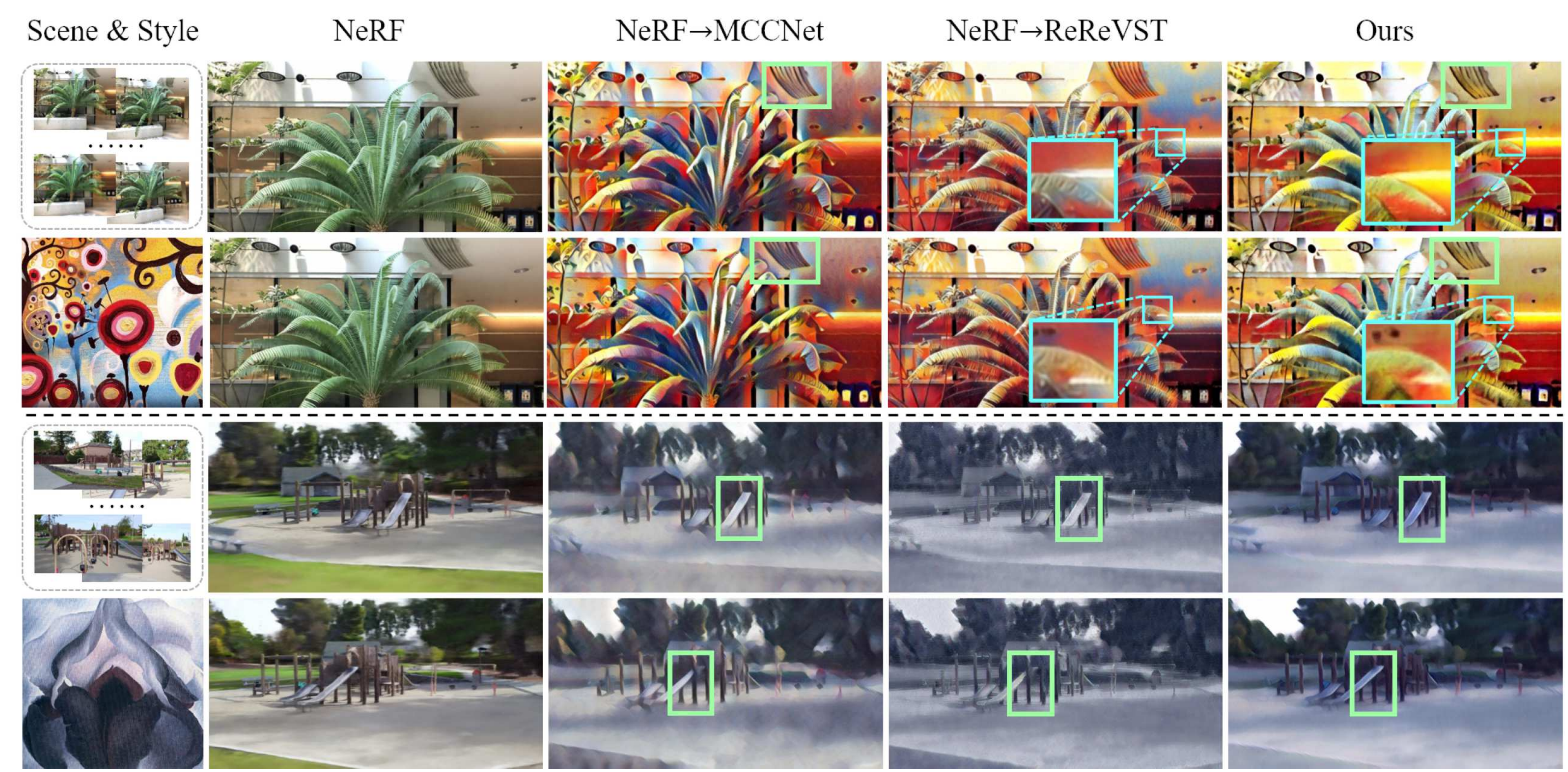}
	\caption{\textbf{Qualitative comparisons to video stylization methods.} We compare the stylized results on novel view videos generated by NeRF on LLFF ~\cite{mildenhall2019local} and Tanks and Temples dataset~\cite{knapitsch2017tanks}. There exist  long-range inconsistencies (1st and 2nd rows) and even appears geometric error (3rd and 4th rows) in the results of video-based approaches.
	}
\label{fig:vs_video}
	\vspace{-4mm}
\end{figure*}

\begin{figure}[htbp]
	\centering
	\includegraphics[width=0.95\linewidth]{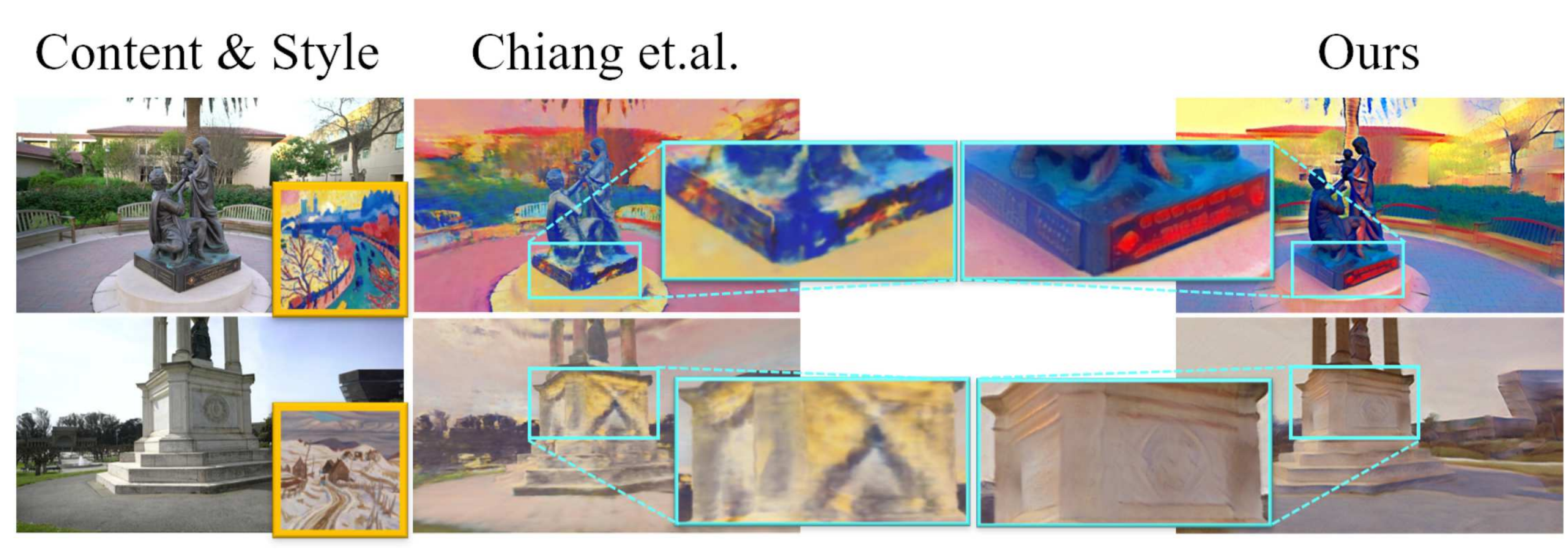}
	\caption{\review{\textbf{Qualitative comparisons to Chiang ~\etal~\cite{chiang2022stylizing}.} We compare the stylized results on Tanks and Temples dataset~\cite{knapitsch2017tanks}. Our results keep better quality of stylized details.
	}
	}
\label{fig:vs_wacv}
	\vspace{-6mm}
\end{figure}

\section{Experiments}
\label{sec:experiment}

We conduct experiments to qualitatively and quantitatively evaluate our method, including comparisons between our method and the state-of-the-art methods of stylization for video and 3D scenes respectively. In quantitative evaluation, a user study is also conducted to collect user preferences, presented in the form of boxplot. We also perform an ablation study on the impact of the ingredients in our method and the effect of training procedure. 
The hyper parameters of $\lambda_d$, $\lambda_s$ and $\lambda_m$ are set to 1e-5, 1 and 10 respectively. 
The style module, latent codes and CNN-based decoder are collaboratively trained for 50k iterations on a single RTX 2080 Ti GPU. The decoder of the 2D stylization method is pre-trained with the consistency loss (Eq.~\ref{eq:consistency}) before the mutual learning process for 1k iterations and fixed in the first 20k iterations of the following collaborative training. We test our method on two types of datasets: forward-facing~\cite{mildenhall2019local} and 360\degree unbounded Tanks \& Templates (T\&T) datasets~\cite{knapitsch2017tanks}.

\subsection{Qualitative Results}
    
    \noindent\textbf{LSNV.} In Fig.~\ref{fig:vs_lsnv}, we qualitatively compare the stylized results of novel views generated by LSNV\cite{huang2021learning} and our method. The geometry representation of LSNV comes from the voxelized point clouds of COLMAP Structure from Motion (SfM) reconstruction~\cite{schonberger2016structure}. The discrete representation results in the absence of fine geometry and the loss of precision, which further damages the stylization results. As is shown in framed yellow boxes, the fine-level shapes like the irregular walls, slender poles, thin chain, etc. are broken and the cracks on the truck are lost and filled in the geometry proxy of LSNV. 
    In contrast, our method gives competitive results thanks to its better preservation of geometry. 
    
    \noindent\textbf{Video stylization.} In Fig.~\ref{fig:vs_video}, we compare our results in multiple views with two state-of-the-art video stylization methods MCCNet~\cite{deng:2020:arbitrary} and ReReVST~\cite{wang2020consistent}.
    Due to the lack of space awareness, video stylization methods cannot guarantee the long-term consistency and even violate the geometry of the scene. Results of the fern in the $1${st}~$2${nd} rows give the examples of long-term inconsistency of 2D methods, where stylized color of the other two methods in the framed areas changes obviously across long-term views. In the results shown in the $3${rd}~$4${th} rows, the geometry of the slide is broken and fused with the pillar behind. 
    Compared with video stylization methods, we conclude that our results are more visually consistent.
    
    \review{
    \noindent\textbf{NeRF-based Stylization.} In Fig.~\ref{fig:vs_wacv}, we compare our results with ~\cite{chiang2022stylizing}, a pioneering work introducing NeRF into stylization. ~\cite{chiang2022stylizing} calculates the style and content losses on small sub-sampled patches which approximate large patches. However, such approximation degrades the preservation of content details, as shown in our comparisons.  Our method produces results with better detail preservation and less artifacts, due to fundamental technical improvements.
    }

\subsection{Quantitative Results}

\noindent\textbf{Consistency Measurement.} Following the measurement in ~\cite{huang2021learning}, we measure the short and long term consistency using the warped LPIPS metric~\cite{zhang2018unreasonable}. A view $v$ is warped with the depth expectation estimated by NeRF.
The score is formulated as:
\begin{equation}
    E(O_i, O_j) = LPIPS(O_i, M_{i,j}, W_{i,j}(O_j))
\end{equation}
where $W$ is the warping function and $M$ is the warping mask. When calculating the average distance across spatial dimensions in ~\cite{zhang2018unreasonable}, only pixels within the mask $M_{i,j}$ are taken. We compute the evaluation values on 4 scenes in the T\&T dataset, using 20 pairs of views for each scene. For each pair, we stylize the images with 10 style images respectively, thus achieve 200 data pairs to evaluate in total. 
The test views are upsampled three times of the training views to ensure the density of frames for video-based methods. 
We use view pairs of gap 5($O_{i},O_{i+5}$) and 35 ($O_{i},O_{i+35}$) for short and long-range consistency calculation. The comparisons of short and long-range consistency are shown in Tab.~\ref{tab:short} and Tab.~\ref{tab:long}, respectively. Our method outperforms other methods by a significant margin.

\begin{table}
\centering
\caption{\textbf{Short-range consistency.} We compare the short-range consistency using warping error($\downarrow$).  \color{RoyalBlue}{\textbf{Best}} \color{black}{and} \color{CornflowerBlue}{second best} \color{black}{results are highlighted.}}\label{tab:short}
\resizebox{\columnwidth}{!}{
\begin{tabular}{l|cccc|c}
\hline
Methods                              & M60   & Truck & Playground & Train & Average  \\ \hline
NeRF $\rightarrow$ AdaIN             & 0.234 & 0.445 & 0.184      & 0.308 & 0.293  \\ \hline
NeRF $\rightarrow$ MCCNet            & 0.169 & 0.167 & 0.134      & 0.261 & 0.183 \\
NeRF $\rightarrow$ ReReVST           & \color{CornflowerBlue}{0.138} & 0.141 & 0.109      & 0.248 & 0.159 \\ \hline
LSNV                                 & 0.139 & \color{CornflowerBlue}{0.138} & \color{CornflowerBlue}{0.104}      & \color{CornflowerBlue}{0.151} & \color{CornflowerBlue}{0.133} \\ \hline

Ours                                 & \color{RoyalBlue}{\textbf{0.063}} & \color{RoyalBlue}{\textbf{0.069}} & \color{RoyalBlue}{\textbf{0.048}} & \color{RoyalBlue}{\textbf{0.115}} & \color{RoyalBlue}{\textbf{0.074}} 
\\ 
\hline
\end{tabular}
}
	\vspace{-2mm}
\end{table}

\begin{table}
\centering
\caption{\textbf{Long-range consistency.} We compare the long-range consistency using warping error($\downarrow$).  \color{RoyalBlue}{\textbf{Best}} \color{black}{and} \color{CornflowerBlue}{second best} \color{black}{results are highlighted.}}\label{tab:long}
\resizebox{\columnwidth}{!}{
\begin{tabular}{l|cccc|c}
\hline
Methods                              & M60   & Truck & Playground & Train & Average  \\ \hline
NeRF $\rightarrow$ AdaIN             & 0.355 & 0.603 & 0.346      & 0.594 & 0.474  \\ \hline
NeRF $\rightarrow$ MCCNet            & 0.307 & 0.317 & 0.290      & 0.567 & 0.370 \\
NeRF $\rightarrow$ ReReVST           & 0.255 & 0.265 & 0.249      & 0.579 & 0.337 \\
\hline
LSNV                                 & \color{CornflowerBlue}{0.206} & \color{CornflowerBlue}{0.239} & \color{CornflowerBlue}{0.193}      & \color{CornflowerBlue}{0.459} & \color{CornflowerBlue}{0.274} \\
\hline
Ours                                 & \color{RoyalBlue}{\textbf{0.139}} & \color{RoyalBlue}{\textbf{0.126}} & \color{RoyalBlue}{\textbf{0.118}} & \color{RoyalBlue}{\textbf{0.205}} & \color{RoyalBlue}{\textbf{0.147}} \\ \hline
\end{tabular}}
	\vspace{-5mm}
\end{table}

\begin{figure}[htbp]
	\vspace{-1mm}
	\centering
	\includegraphics[width=0.95\linewidth]{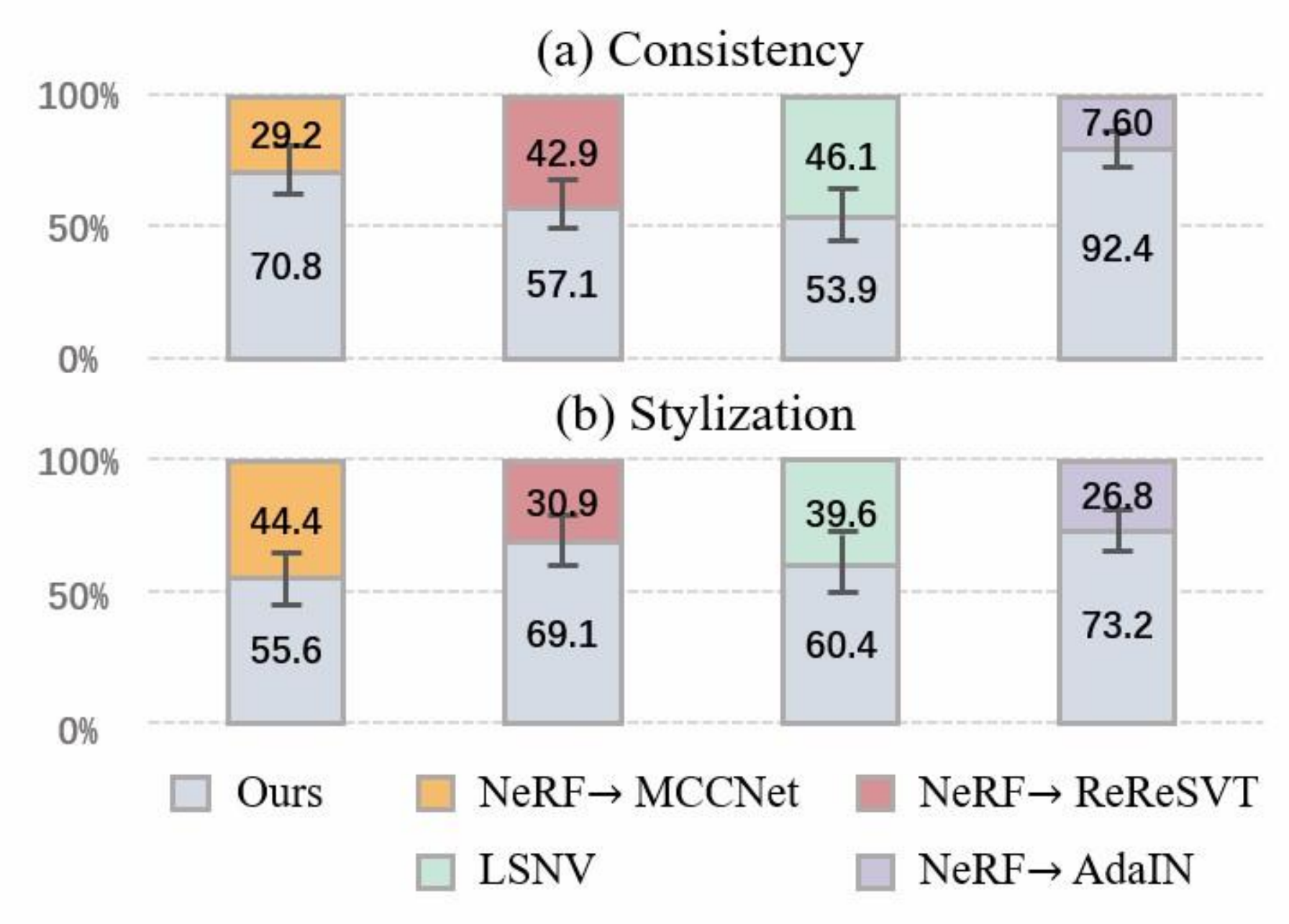}
	\caption{\textbf{User study.} We record the user preference in the form of boxplot. Our results win more preferences both in the stylization and consistency quality.}
\label{fig:user_study}
	\vspace{-4mm}
\end{figure}

\noindent\textbf{User study.} A user study is conducted to compare the stylization and consistency quality of our method with other state-of-the-art methods. We stylize ten series of views of the 3D scenes in the T\&T dataset, using different methods~\cite{deng:2020:arbitrary},~\cite{wang2020consistent},~\cite{huang2021learning} and invite 50 participants (including 28 males, 22 females, aged from 18 to 45). First we showed the participants a style image and two stylized videos generated by our method and a random compared method. Then we asked the participants their votes for the video in two evaluating indicators, quality of the stylized results and whether to keep the consistency. We collected 1000 votes for each evaluating indicator and present the result in Fig.~\ref{fig:user_study} in the form of boxplot. Our scores stand out from other methods in both stylization quality and consistency.

\subsection{Ablation Study}

\begin{figure}[htbp]
	\vspace{-1mm}
	\centering
	\includegraphics[width=0.95\linewidth]{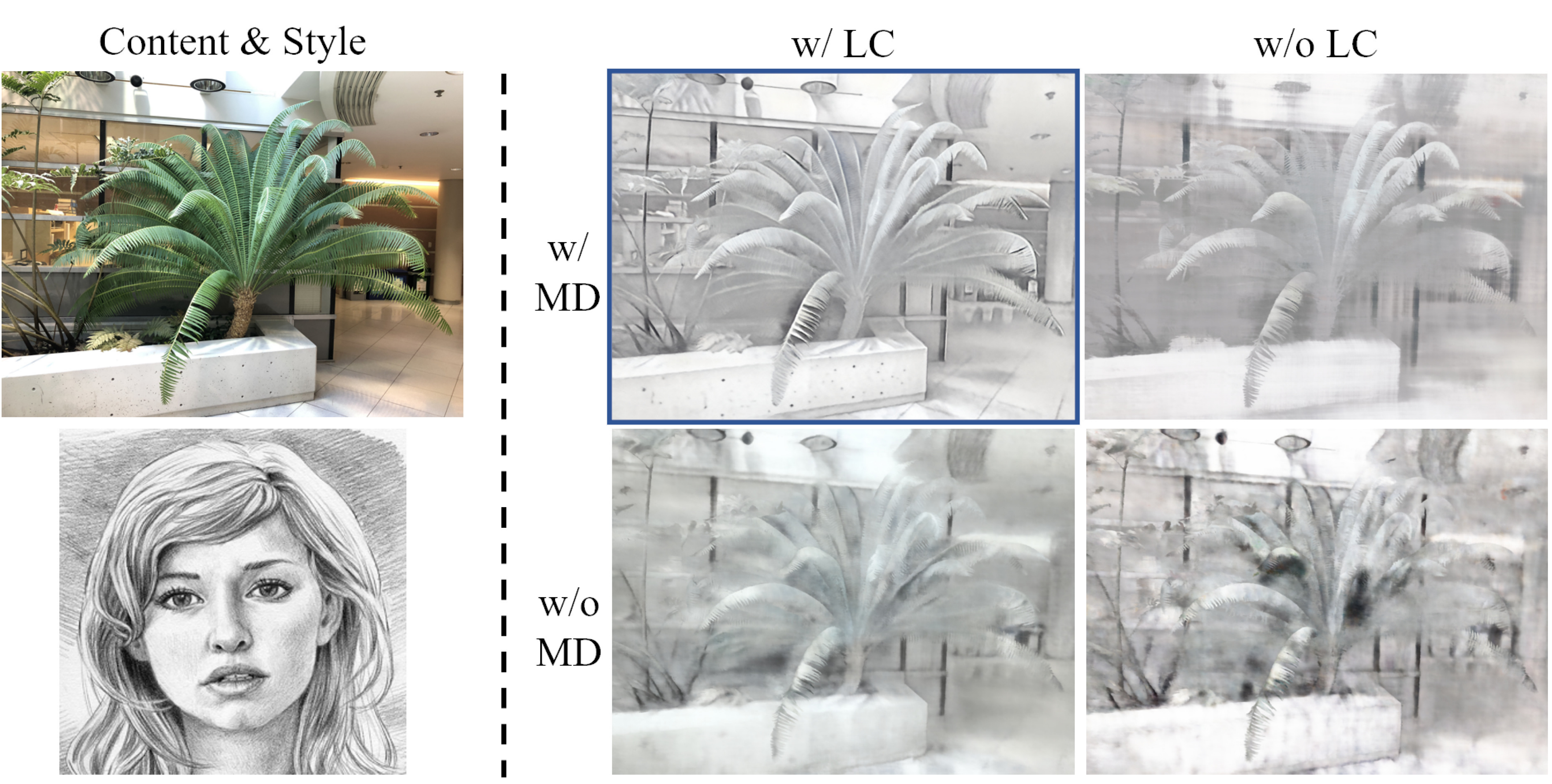}
	\caption{\textbf{The impact of the learnable codes design (LC) and mutual learning the decoder (MD).} Mutually training 2D method and applying learnable latent codes both raise the quality of stylized results.}
\label{fig:variance}
	\vspace{-4mm}
\end{figure}

\noindent\textbf{The impact of the learnable codes design and mutual training scheme with learnable 2D method.} Compared to the choice of shared and fixed style latent codes (w/o LC), applying learnable latent codes (w/ LC) helps to handle the inconsistency of distilled knowledge $\mathcal{C}_a(r,\mathcal(S))$ from the 2D method. On the other hand, to train the decoder (MD) in the mutual learning process is another operation we take to raise the consistency level of 2D method's outputs and thus to make it easier to train the latent codes.
It can be obviously seen that without one or both of these designs mentioned above, the artifacts and blurriness appear in the results as shown in Fig.~\ref{fig:variance}. It demonstrates the necessity and robustness of our design, whose result is of clearer object outlines and more reasonable stylization.

\begin{figure}[htbp]
 	\centering
 	\includegraphics[width=0.95\linewidth]{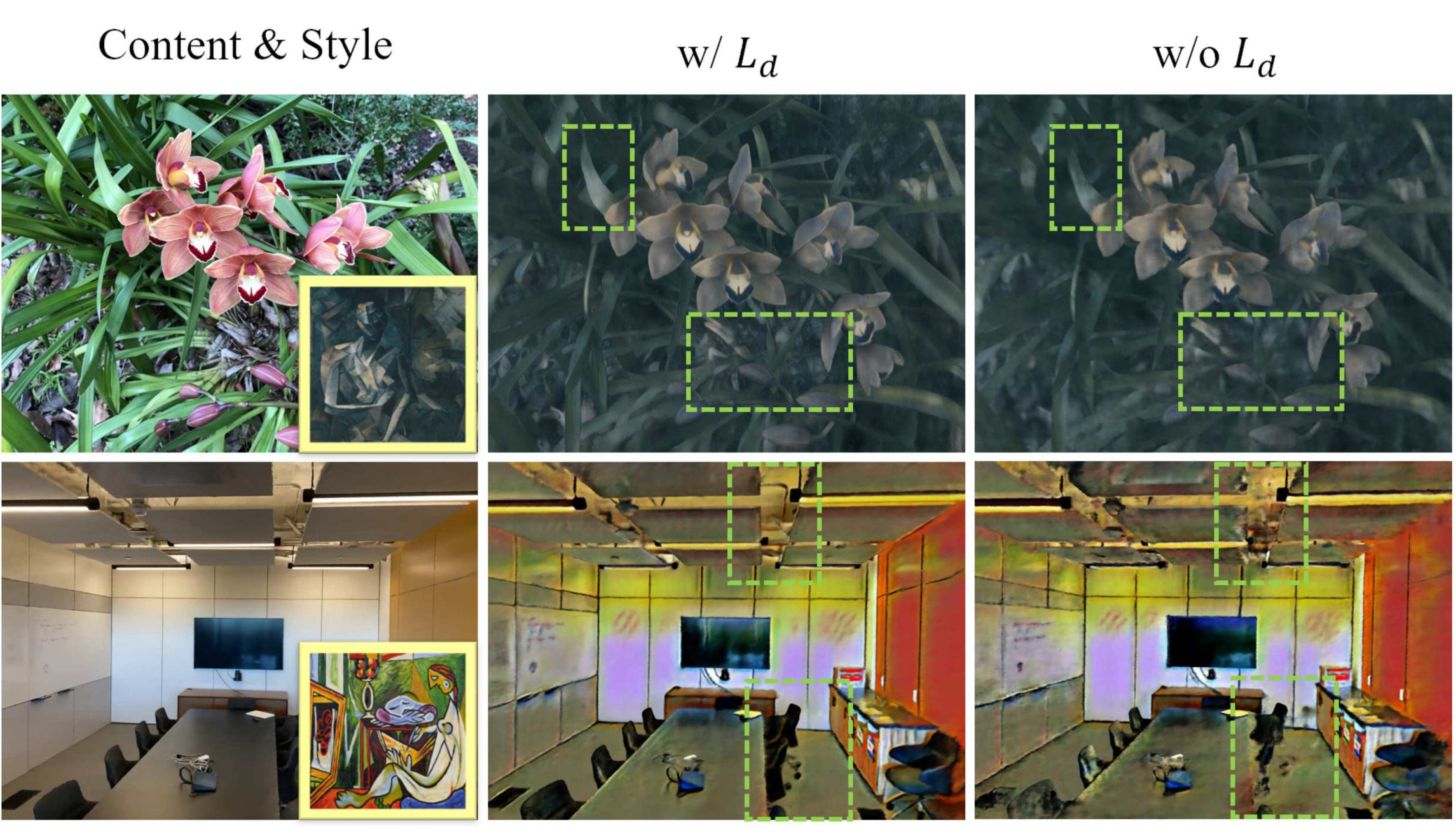}
 	\caption{\textbf{Ablation study on the impact of $L_d$.} $L_d$ clusters latent codes of the same style and avoids the artifacts in test results.}
\label{fig:logp}
 	\vspace{-5mm}
\end{figure}

\noindent\textbf{The impact of $L_d$. }  At the inference time, the mean code $\mu$ of the encoded distribution $\mathcal{N}(\mu, \sigma)$ is used as input to the style module of NeRF.
Fig.~\ref{fig:logp} compares the inference results with and without $L_d$. We use green boxes to frame obvious artifacts in the results without $L_d$ in the third column, while results of our complete network in the second column handles this well. This distribution loss constrains the learnable latent codes to obtain better clustering around the mean of the pre-trained distribution, and helps avoid artifacts during inference process.

\begin{figure}[htbp]
 	\centering
 	\includegraphics[width=1.\linewidth]{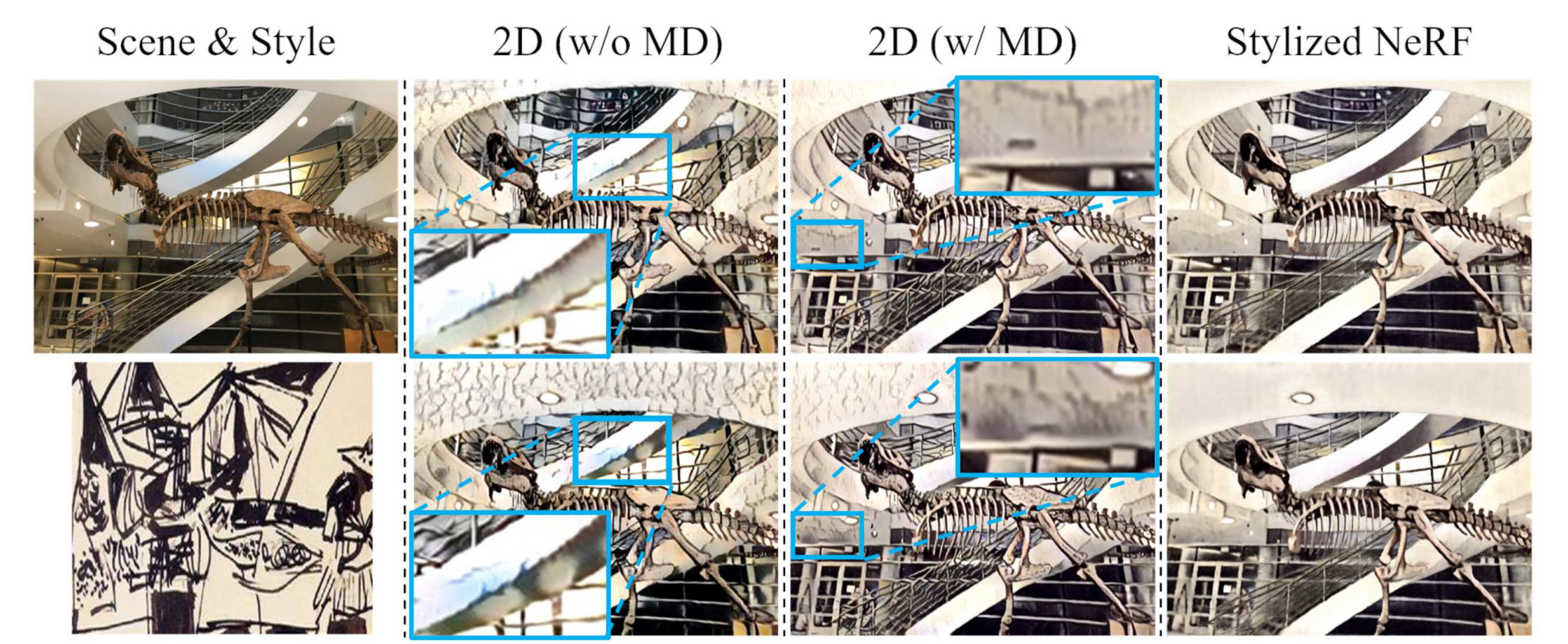}
 	\caption{\textbf{Ablation study on the choice of results.} We compare the results of the stylized NeRF with the results of 2D method with (MD) and without (w/o MD) mutual learning. The results of stylized NeRF keeps the best consistency.}
\label{fig:2d}
 	\vspace{-4mm}
\end{figure}

\noindent\textbf{The choice of outputs.} Our method produces two results of 2D method and stylized NeRF at every iteration in a mutual-learning way. We compare the two obtained results as shown in the Fig.~\ref{fig:2d}. Although the mutual learning process makes the 2D method ($2${nd} and $3${rd} columns) iterates towards a more consistent trend, its consistency is not strict enough and encounters flickering issues between long-range views. On the contrary, the rendered results of stylized NeRF ($4${th} column) keep the excellent consistency thanks to its physical volume rendering scheme. Hence we choose the outputs of stylized NeRF as our final results.

\section{Conclusion}
\hyh{We present StylizedNeRF, a novel method for stylizing 3D scenes. A novel mutual-learning framework is proposed to best leverage the stylization ability of 2D method and the spatial consistency of NeRF. 
A consistency loss distilling spatial consistency prior from NeRF to 2D networks and a mimic loss aligning outputs of 2D networks and stylized NeRF are introduced. 
To further suppress the inconsistency of 2D method and enable the conditional stylization, we parameterize the inconsistent 2D stylized results as latent codes obeying the distributions conditioned on styles. Our StylizedNeRF outperforms state-of-the-art methods both in terms of visual quality and  consistency. In future work, we will implement our
proposed approach in Jittor~\cite{hu2020jittor}, which is a fully just-in-time (JIT) compiled deep learning framework.}

\section*{Acknowledgement}
\noindent This work was supported by the Beijing Municipal Natural Science Foundation for Distinguished Young Scholars (No. JQ21013), the National Natural Science Foundation of China (No. 62061136007 and No. 61872440), Royal Society Newton Advanced Fellowship (No. NAF\verb|\|R2\verb|\|192151) and the Youth Innovation Promotion Association CAS.

{\small
\bibliographystyle{ieee_fullname}
\bibliography{egbib}

\begin{thebibliography}{10}\itemsep=-1pt

\bibitem{aliev2020neural}
Kara-Ali Aliev, Artem Sevastopolsky, Maria Kolos, Dmitry Ulyanov, and Victor
  Lempitsky.
\newblock Neural point-based graphics.
\newblock In {\em Proceedings of the European Conference on Computer Vision
  (ECCV)}, pages 696--712. Springer, 2020.

\bibitem{boss2021nerd}
Mark Boss, Raphael Braun, Varun Jampani, Jonathan~T Barron, Ce Liu, and Hendrik
  Lensch.
\newblock {NeRD}: Neural reflectance decomposition from image collections.
\newblock In {\em Proceedings of the IEEE/CVF International Conference on
  Computer Vision (ICCV)}, pages 12684--12694, 2021.

\bibitem{buehler2001unstructured}
Chris Buehler, Michael Bosse, Leonard McMillan, Steven Gortler, and Michael
  Cohen.
\newblock Unstructured lumigraph rendering.
\newblock In {\em Proceedings of the 28th annual conference on Computer
  graphics and interactive techniques}, pages 425--432, 2001.

\bibitem{chen2017coherent}
Dongdong Chen, Jing Liao, Lu Yuan, Nenghai Yu, and Gang Hua.
\newblock Coherent online video style transfer.
\newblock In {\em Proceedings of the IEEE/CVF International Conference on
  Computer Vision (ICCV)}, pages 1105--1114, 2017.

\bibitem{chiang2022stylizing}
Pei-Ze Chiang, Meng-Shiun Tsai, Hung-Yu Tseng, Wei-Sheng Lai, and Wei-Chen
  Chiu.
\newblock Stylizing {3D} scene via implicit representation and {HyperNetwork}.
\newblock In {\em Proceedings of the IEEE/CVF Winter Conference on Applications
  of Computer Vision (WACV)}, pages 1475--1484, 2022.

\bibitem{davis2012unstructured}
Abe Davis, Marc Levoy, and Fredo Durand.
\newblock Unstructured light fields.
\newblock {\em Computer Graphics Forum}, 31(2pt1):305--314, 2012.

\bibitem{debevec1996modeling}
Paul~E Debevec, Camillo~J Taylor, and Jitendra Malik.
\newblock Modeling and rendering architecture from photographs: A hybrid
  geometry-and image-based approach.
\newblock In {\em Proceedings of the 23rd annual conference on Computer
  graphics and interactive techniques}, pages 11--20, 1996.

\bibitem{deng:2020:arbitrary}
Yingying Deng, Fan Tang, Weiming Dong, haibin Huang, Ma chongyang, and
  Changsheng Xu.
\newblock Arbitrary video style transfer via multi-channel correlation.
\newblock In {\em AAAI}, 2021.

\bibitem{flynn2019deepview}
John Flynn, Michael Broxton, Paul Debevec, Matthew DuVall, Graham Fyffe, Ryan
  Overbeck, Noah Snavely, and Richard Tucker.
\newblock {DeepView}: View synthesis with learned gradient descent.
\newblock In {\em Proceedings of the IEEE/CVF Conference on Computer Vision and
  Pattern Recognition (CVPR)}, pages 2367--2376, 2019.

\bibitem{gao2020tmnet}
Lin Gao, Tong Wu, Yu-Jie Yuan, Ming-Xian Lin, Yu-Kun Lai, and Hao Zhang.
\newblock {TM-NET}: Deep generative networks for textured meshes.
\newblock {\em ACM Transactions on Graphics (TOG)}, 40(6):263:1--263:15, 2021.

\bibitem{gao2018automatic}
Lin Gao, Jie Yang, Yi-Ling Qiao, Yu-Kun Lai, Paul~L Rosin, Weiwei Xu, and
  Shihong Xia.
\newblock Automatic unpaired shape deformation transfer.
\newblock {\em ACM Transactions on Graphics (TOG)}, 37(6):1--15, 2018.

\bibitem{gao2020fast}
Wei Gao, Yijun Li, Yihang Yin, and Ming-Hsuan Yang.
\newblock Fast video multi-style transfer.
\newblock In {\em Proceedings of the IEEE/CVF Winter Conference on Applications
  of Computer Vision (WACV)}, pages 3222--3230, 2020.

\bibitem{gatys2015neural}
Leon~A Gatys, Alexander~S Ecker, and Matthias Bethge.
\newblock A neural algorithm of artistic style.
\newblock {\em Nature Communications}, 2015.

\bibitem{Gatys2015TextureSU}
Leon~A. Gatys, Alexander~S. Ecker, and M. Bethge.
\newblock Texture synthesis using convolutional neural networks.
\newblock In {\em Advances in Neural Information Processing Systems (NeurIPS)},
  pages 262--270, 2015.

\bibitem{gortler1996lumigraph}
Steven~J Gortler, Radek Grzeszczuk, Richard Szeliski, and Michael~F Cohen.
\newblock The lumigraph.
\newblock In {\em Proceedings of the 23rd annual conference on Computer
  graphics and interactive techniques}, pages 43--54, 1996.

\bibitem{habtegebrial2020generative}
Tewodros Habtegebrial, Varun Jampani, Orazio Gallo, and Didier Stricker.
\newblock Generative view synthesis: From single-view semantics to novel-view
  images.
\newblock In {\em Advances in Neural Information Processing Systems (NeurIPS)},
  2020.

\bibitem{hu2020jittor}
Shi-Min Hu, Dun Liang, Guo-Ye Yang, Guo-Wei Yang, and Wen-Yang Zhou.
\newblock Jittor: a novel deep learning framework with meta-operators and
  unified graph execution.
\newblock {\em Science China Information Sciences}, 63(222103):1--21, 2020.

\bibitem{huang2020semantic}
Hsin-Ping Huang, Hung-Yu Tseng, Hsin-Ying Lee, and Jia-Bin Huang.
\newblock Semantic view synthesis.
\newblock In {\em Proceedings of the European Conference on Computer Vision
  (ECCV)}, pages 592--608. Springer, 2020.

\bibitem{huang2021learning}
Hsin-Ping Huang, Hung-Yu Tseng, Saurabh Saini, Maneesh Singh, and Ming-Hsuan
  Yang.
\newblock Learning to stylize novel views.
\newblock In {\em Proceedings of the IEEE/CVF International Conference on
  Computer Vision (ICCV)}, pages 13869--13878, 2021.

\bibitem{huang2017arbitrary}
Xun Huang and Serge Belongie.
\newblock Arbitrary style transfer in real-time with adaptive instance
  normalization.
\newblock In {\em Proceedings of the IEEE/CVF International Conference on
  Computer Vision (ICCV)}, pages 1501--1510, 2017.

\bibitem{johnson2016perceptual}
Justin Johnson, Alexandre Alahi, and Li Fei-Fei.
\newblock Perceptual losses for real-time style transfer and super-resolution.
\newblock In {\em Proceedings of the European conference on computer vision
  (ECCV)}, pages 694--711. Springer, 2016.

\bibitem{kanazawa2018learning}
Angjoo Kanazawa, Shubham Tulsiani, Alexei~A Efros, and Jitendra Malik.
\newblock Learning category-specific mesh reconstruction from image
  collections.
\newblock In {\em Proceedings of the European Conference on Computer Vision
  (ECCV)}, pages 371--386, 2018.

\bibitem{kingma2013auto}
Diederik~P Kingma and Max Welling.
\newblock Auto-encoding variational bayes.
\newblock In {\em ICLR}, 2014.

\bibitem{knapitsch2017tanks}
Arno Knapitsch, Jaesik Park, Qian-Yi Zhou, and Vladlen Koltun.
\newblock Tanks and temples: Benchmarking large-scale scene reconstruction.
\newblock {\em ACM Transactions on Graphics (TOG)}, 36(4):1--13, 2017.

\bibitem{kotovenko2021rethinking}
Dmytro Kotovenko, Matthias Wright, Arthur Heimbrecht, and Bjorn Ommer.
\newblock Rethinking style transfer: From pixels to parameterized brushstrokes.
\newblock In {\em Proceedings of the IEEE/CVF Conference on Computer Vision and
  Pattern Recognition (CVPR)}, pages 12196--12205, 2021.

\bibitem{levoy1996light}
Marc Levoy and Pat Hanrahan.
\newblock Light field rendering.
\newblock In {\em Proceedings of the 23rd annual conference on Computer
  graphics and interactive techniques}, pages 31--42, 1996.

\bibitem{Univeral_style_transfer}
Yijun Li, Chen Fang, Jimei Yang, Zhaowen Wang, Xin Lu, and Ming{-}Hsuan Yang.
\newblock Universal style transfer via feature transforms.
\newblock In {\em Advances in Neural Information Processing Systems (NeurIPS)},
  pages 386--396, 2017.

\bibitem{li2018closed}
Yijun Li, Ming-Yu Liu, Xueting Li, Ming-Hsuan Yang, and Jan Kautz.
\newblock A closed-form solution to photorealistic image stylization.
\newblock In {\em Proceedings of the European Conference on Computer Vision
  (ECCV)}, pages 453--468, 2018.

\bibitem{liu2020neural}
Lingjie Liu, Jiatao Gu, Kyaw~Zaw Lin, Tat-Seng Chua, and Christian Theobalt.
\newblock Neural sparse voxel fields.
\newblock In {\em Advances in Neural Information Processing Systems (NeurIPS)},
  2020.

\bibitem{lombardi2019neural}
Stephen Lombardi, Tomas Simon, Jason Saragih, Gabriel Schwartz, Andreas
  Lehrmann, and Yaser Sheikh.
\newblock Neural volumes: Learning dynamic renderable volumes from images.
\newblock {\em ACM Transactions on Graphics (TOG)}, 2019.

\bibitem{martin2021nerf}
Ricardo Martin-Brualla, Noha Radwan, Mehdi~SM Sajjadi, Jonathan~T Barron,
  Alexey Dosovitskiy, and Daniel Duckworth.
\newblock {NeRF} in the wild: Neural radiance fields for unconstrained photo
  collections.
\newblock In {\em Proceedings of the IEEE/CVF Conference on Computer Vision and
  Pattern Recognition (CVPR)}, pages 7210--7219, 2021.

\bibitem{max1995optical}
Nelson Max.
\newblock Optical models for direct volume rendering.
\newblock {\em IEEE Transactions on Visualization and Computer Graphics
  (TVCG)}, 1(2):99--108, 1995.

\bibitem{meshry2019neural}
Moustafa Meshry, Dan~B Goldman, Sameh Khamis, Hugues Hoppe, Rohit Pandey, Noah
  Snavely, and Ricardo Martin-Brualla.
\newblock Neural rerendering in the wild.
\newblock In {\em Proceedings of the IEEE/CVF Conference on Computer Vision and
  Pattern Recognition (CVPR)}, pages 6878--6887, 2019.

\bibitem{mildenhall2019local}
Ben Mildenhall, Pratul~P Srinivasan, Rodrigo Ortiz-Cayon, Nima~Khademi
  Kalantari, Ravi Ramamoorthi, Ren Ng, and Abhishek Kar.
\newblock Local light field fusion: Practical view synthesis with prescriptive
  sampling guidelines.
\newblock {\em ACM Transactions on Graphics (TOG)}, 38(4):1--14, 2019.

\bibitem{mildenhall2020nerf}
Ben Mildenhall, Pratul~P Srinivasan, Matthew Tancik, Jonathan~T Barron, Ravi
  Ramamoorthi, and Ren Ng.
\newblock {NeRF}: Representing scenes as neural radiance fields for view
  synthesis.
\newblock In {\em Proceedings of the European conference on computer vision
  (ECCV)}, pages 405--421. Springer, 2020.

\bibitem{niemeyer2020differentiable}
Michael Niemeyer, Lars Mescheder, Michael Oechsle, and Andreas Geiger.
\newblock Differentiable volumetric rendering: Learning implicit {3D}
  representations without {3D} supervision.
\newblock In {\em Proceedings of the IEEE/CVF Conference on Computer Vision and
  Pattern Recognition (CVPR)}, pages 3504--3515, 2020.

\bibitem{niklaus20193d}
Simon Niklaus, Long Mai, Jimei Yang, and Feng Liu.
\newblock {3D} {Ken} {Burns} effect from a single image.
\newblock {\em ACM Transactions on Graphics (TOG)}, 38(6):1--15, 2019.

\bibitem{riegler2020free}
Gernot Riegler and Vladlen Koltun.
\newblock Free view synthesis.
\newblock In {\em Proceedings of the European Conference on Computer Vision
  (ECCV)}, pages 623--640. Springer, 2020.

\bibitem{riegler2021stable}
Gernot Riegler and Vladlen Koltun.
\newblock Stable view synthesis.
\newblock In {\em Proceedings of the IEEE/CVF Conference on Computer Vision and
  Pattern Recognition (CVPR)}, pages 12216--12225, 2021.

\bibitem{schonberger2016structure}
Johannes~L Schonberger and Jan-Michael Frahm.
\newblock Structure-from-motion revisited.
\newblock In {\em Proceedings of the IEEE/CVF conference on computer vision and
  pattern recognition (CVPR)}, pages 4104--4113, 2016.

\bibitem{seitz1999photorealistic}
Steven~M Seitz and Charles~R Dyer.
\newblock Photorealistic scene reconstruction by voxel coloring.
\newblock {\em International Journal of Computer Vision}, 35(2):151--173, 1999.

\bibitem{sheng2018avatar}
Lu Sheng, Ziyi Lin, Jing Shao, and Xiaogang Wang.
\newblock Avatar-net: Multi-scale zero-shot style transfer by feature
  decoration.
\newblock In {\em Proceedings of the IEEE/CVF Conference on Computer Vision and
  Pattern Recognition (CVPR)}, pages 8242--8250, 2018.

\bibitem{simonyan2014very}
Karen Simonyan and Andrew Zisserman.
\newblock Very deep convolutional networks for large-scale image recognition.
\newblock In {\em ICLR}, 2015.

\bibitem{sitzmann2019deepvoxels}
Vincent Sitzmann, Justus Thies, Felix Heide, Matthias Nie{\ss}ner, Gordon
  Wetzstein, and Michael Zollhofer.
\newblock {DeepVoxels}: Learning persistent {3D} feature embeddings.
\newblock In {\em Proceedings of the IEEE/CVF Conference on Computer Vision and
  Pattern Recognition (CVPR)}, pages 2437--2446, 2019.

\bibitem{srinivasan2019pushing}
Pratul~P Srinivasan, Richard Tucker, Jonathan~T Barron, Ravi Ramamoorthi, Ren
  Ng, and Noah Snavely.
\newblock Pushing the boundaries of view extrapolation with multiplane images.
\newblock In {\em Proceedings of the IEEE/CVF Conference on Computer Vision and
  Pattern Recognition (CVPR)}, pages 175--184, 2019.

\bibitem{sumner2004deformation}
Robert~W Sumner and Jovan Popovi{\'c}.
\newblock Deformation transfer for triangle meshes.
\newblock {\em ACM Transactions on graphics (TOG)}, 23(3):399--405, 2004.

\bibitem{tancik2020fourier}
Matthew Tancik, Pratul Srinivasan, Ben Mildenhall, Sara Fridovich-Keil, Nithin
  Raghavan, Utkarsh Singhal, Ravi Ramamoorthi, Jonathan Barron, and Ren Ng.
\newblock Fourier features let networks learn high frequency functions in low
  dimensional domains.
\newblock In {\em Advances in Neural Information Processing Systems (NeurIPS)},
  volume~33, pages 7537--7547, 2020.

\bibitem{tewari2021advances}
Ayush Tewari, O Fried, J Thies, V Sitzmann, S Lombardi, Z Xu, T Simon, M
  Nie{\ss}ner, E Tretschk, L Liu, et~al.
\newblock Advances in neural rendering.
\newblock In {\em ACM SIGGRAPH 2021 Courses}, pages 1--320. 2021.

\bibitem{tucker2020single}
Richard Tucker and Noah Snavely.
\newblock Single-view view synthesis with multiplane images.
\newblock In {\em Proceedings of the IEEE/CVF Conference on Computer Vision and
  Pattern Recognition (CVPR)}, pages 551--560, 2020.

\bibitem{waechter2014let}
Michael Waechter, Nils Moehrle, and Michael Goesele.
\newblock Let there be color! large-scale texturing of {3D} reconstructions.
\newblock In {\em Proceedings of the European conference on computer vision
  (ECCV)}, pages 836--850. Springer, 2014.

\bibitem{wang2020consistent}
Wenjing Wang, Shuai Yang, Jizheng Xu, and Jiaying Liu.
\newblock Consistent video style transfer via relaxation and regularization.
\newblock {\em IEEE Transactions on Image Processing}, 29:9125--9139, 2020.

\bibitem{wiles2020synsin}
Olivia Wiles, Georgia Gkioxari, Richard Szeliski, and Justin Johnson.
\newblock {SynSin}: End-to-end view synthesis from a single image.
\newblock In {\em Proceedings of the IEEE/CVF Conference on Computer Vision and
  Pattern Recognition (CVPR)}, pages 7467--7477, 2020.

\bibitem{wizadwongsa2021nex}
Suttisak Wizadwongsa, Pakkapon Phongthawee, Jiraphon Yenphraphai, and Supasorn
  Suwajanakorn.
\newblock {NeX}: Real-time view synthesis with neural basis expansion.
\newblock In {\em Proceedings of the IEEE/CVF Conference on Computer Vision and
  Pattern Recognition (CVPR)}, pages 8534--8543, 2021.

\bibitem{wood2000surface}
Daniel~N Wood, Daniel~I Azuma, Ken Aldinger, Brian Curless, Tom Duchamp,
  David~H Salesin, and Werner Stuetzle.
\newblock Surface light fields for {3D} photography.
\newblock In {\em Proceedings of the 27th annual conference on Computer
  graphics and interactive techniques}, pages 287--296, 2000.

\bibitem{xiang2021neutex}
Fanbo Xiang, Zexiang Xu, Milos Hasan, Yannick Hold-Geoffroy, Kalyan Sunkavalli,
  and Hao Su.
\newblock {NeuTex}: Neural texture mapping for volumetric neural rendering.
\newblock In {\em Proceedings of the IEEE/CVF Conference on Computer Vision and
  Pattern Recognition (CVPR)}, pages 7119--7128, 2021.

\bibitem{yariv2020multiview}
Lior Yariv, Yoni Kasten, Dror Moran, Meirav Galun, Matan Atzmon, Basri Ronen,
  and Yaron Lipman.
\newblock Multiview neural surface reconstruction by disentangling geometry and
  appearance.
\newblock In {\em Advances in Neural Information Processing Systems (NeurIPS)},
  volume~33, pages 2492--2502, 2020.

\bibitem{yoo2019photorealistic}
Jaejun Yoo, Youngjung Uh, Sanghyuk Chun, Byeongkyu Kang, and Jung-Woo Ha.
\newblock Photorealistic style transfer via wavelet transforms.
\newblock In {\em Proceedings of the IEEE/CVF International Conference on
  Computer Vision (ICCV)}, pages 9036--9045, 2019.

\bibitem{zhang2020nerf++}
Kai Zhang, Gernot Riegler, Noah Snavely, and Vladlen Koltun.
\newblock {NeRF}++: Analyzing and improving neural radiance fields.
\newblock {\em arXiv preprint arXiv:2010.07492}, 2020.

\bibitem{zhang2018unreasonable}
Richard Zhang, Phillip Isola, Alexei~A Efros, Eli Shechtman, and Oliver Wang.
\newblock The unreasonable effectiveness of deep features as a perceptual
  metric.
\newblock In {\em Proceedings of the IEEE conference on computer vision and
  pattern recognition (CVPR)}, pages 586--595, 2018.

\bibitem{zhang2018deep}
Ying Zhang, Tao Xiang, Timothy~M Hospedales, and Huchuan Lu.
\newblock Deep mutual learning.
\newblock In {\em Proceedings of the IEEE/CVF Conference on Computer Vision and
  Pattern Recognition (CVPR)}, pages 4320--4328, 2018.

\bibitem{zheng2021weakly}
Qian Zheng, Yajie Liu, Zhizhao Lin, Dani Lischinski, Daniel Cohen-Or, and Hui
  Huang.
\newblock Weakly supervised {2D} human pose transfer.
\newblock {\em Science China Information Sciences}, 64(11):1--16, 2021.

\bibitem{zhou2018stereo}
Tinghui Zhou, Richard Tucker, John Flynn, Graham Fyffe, and Noah Snavely.
\newblock Stereo magnification: Learning view synthesis using multiplane
  images.
\newblock {\em ACM Transactions on Graphics (TOG)}, 2018.

\end{thebibliography}
}

\end{document}